\documentclass[twocolumn]{jpsj3}

\usepackage{color}
\usepackage{bm}

\title{%
Persistent Current due to a Screw Dislocation
in Weyl Semimetals: Role of One-Dimensional Chiral States
}

\author{%
Kentaro Kodama and Yositake Takane
}

\inst{%
Department of Quantum Matter, Graduate School of Advanced Sciences of Matter,\\
Hiroshima University, Higashihiroshima, Hiroshima 739-8530, Japan
}

\recdate{ \hspace{50mm} }

\abst{%
A Weyl semimetal pierced by a screw dislocation accommodates
one-dimensional (1D) chiral states along the corresponding dislocation line.
As these states propagate in a particular direction determined by
their chirality, a persistent current (i.e., charge current in equilibrium)
is expected to appear in the interior of the system.
To confirm this expectation, we numerically calculate the charge current
in a Weyl semimetal in the presence of a screw dislocation.
It is shown that a significant charge current is induced by
the 1D chiral states near the dislocation.
We also analyze the spatial distribution of the charge current
focusing on the top and bottom surfaces of the system,
at which the screw dislocation is terminated, and give an overview of
how the charge current due to the dislocation is converted to
that carried by other states near the termination point of the dislocation.
}

\begin{document}
\sloppy
\maketitle

\section{Introduction}

A Weyl semimetal is a three-dimensional topological system
with gapless excitations.
It possesses a pair of, or pairs of, nondegenerate Dirac cones
with opposite chirality.~\cite{shindou,murakami,wan,yang,
burkov1,burkov2,WK,delplace,halasz,sekine}
The conduction and valence bands of each Dirac cone conically touch
at the point called a Weyl node in the Brillouin zone.
A notable feature of a Weyl semimetal is that low-energy chiral states
appear on its surface in a two-dimensional (2D) manner~\cite{wan}
if a pair of Weyl nodes is projected onto two different points
in the corresponding surface Brillouin zone.
These 2D chiral surface states are collectively referred to as a Fermi arc
as they appear to connect a pair of projected Weyl nodes.
It has been shown that similar chiral states also appear in a Weyl semimetal
in the presence of a screw dislocation (Fig.~1)~\cite{imura1,takane1}
or an edge dislocation.~\cite{takane2}
In this case, the chiral states are localized
along each dislocation line as those observed
in topological insulators;~\cite{ran,zhang,imura2,yoshimura,pauly}
thus, they appear in a one-dimensional (1D) manner.
Some materials, such as TaAs and NbAs, have been experimentally identified
as Weyl semimetals.~\cite{weng,huang1,xu1,lv1,lv2,xu2,souma,kuroda}

For definiteness, we consider a prototypical Weyl semimetal
with a pair of Weyl nodes at $\mib{k}_{\pm} = (0,0,\pm k_{0})$
in reciprocal space and $E = 0$ in energy space.
We assume that the system is in the shape of a prism or cylinder
of finite length aligned along the $z$-direction.
In this case, the 2D chiral surface states appear
on the side surface of the system.
Let us focus on the 1D chiral states along a screw dislocation
parallel to the $z$-axis.
As the 1D chiral states propagate only in a particular direction determined
by their chirality, they collectively carry a charge current
in the direction opposite to its propagating direction.
This suggests the possibility that a persistent charge current is
spontaneously induced in the interior of a Weyl semimetal in equilibrium.

\begin{figure}[btp]
\begin{center}
\includegraphics[height=2.5cm]{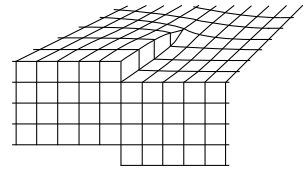}
\end{center}
\caption{
Screw dislocation with a displacement of one unit atomic layer
terminated on a top surface.
}
\end{figure}
Previously, Sumiyoshi and Fujimoto~\cite{sumiyoshi} analyzed
the response of a Weyl semimetal upon the insertion of a screw (or edge)
dislocation to answer the question of whether
the fictitious magnetic field~\cite{kawamura} due to the dislocation
gives rise to a local charge current.
They found that a local charge current does appear along the dislocation.
This phenomenon can be regarded as a chiral magnetic effect
in equilibrium, which is forbidden for a uniform magnetic field.~\cite{zyuzin,
chen,vazifeh,chang1,chang2,yamamoto,takane3,ibe,takane4}
However, they did not analyze the case in which 1D chiral states are
accommodated near a screw dislocation.
That is, the local charge current considered in Ref.~\citen{sumiyoshi}
is induced by the response of bulk states to the fictitious magnetic field.
It is natural to expect that the 1D chiral states more significantly
contribute to the local charge current than the bulk states.
However, this has not been verified yet.
Furthermore, the distribution of charge current has not been explicitly
examined in a realistic system of finite size.
The charge current in equilibrium must vanish if it is integrated over
an entire cross section of singly connected systems.
That is, the charge current near the dislocation is canceled out by
that near the side surface, which should flow in the opposite direction,
as well as by that due to bulk states.
This observation gives rise to the natural question: how is the charge current
near the dislocation converted to the surface charge current
at the top and bottom surfaces?

In this paper, we analyze the local charge current in a Weyl semimetal
to answer the question raised above.
Using a standard tight-binding model for a Weyl semimetal,
which possesses particle-hole symmetry,
we calculate the spatial distribution of the charge current
in the presence of a screw dislocation.
It is shown that the local charge current significantly depends on
the location of the Fermi energy $E_{F}$.
If $E_{F}$ is at the band center,
which corresponds to the energy at the Weyl nodes (i.e., $E = 0$),
the local charge current completely vanishes everywhere
in the system even in the presence of a screw dislocation.
A finite charge current is observed near the dislocation
once $E_{F}$ is displaced from the band center.
It is clearly shown that the contribution to the charge current from
the 1D chiral states significantly dominates that from the bulk states.
Indeed, the local charge current becomes significantly small
in the absence of the 1D chiral states.
It is also shown that the charge current is nearly proportional to $E_{F}$
as long as $E_{F}$ is located near the band center.
We finally examine how the charge current near a screw dislocation is
converted to the surface current at the top and bottom surfaces of the system.
It is shown that the conversion is mainly mediated through
the edge dislocation,
which connects the screw dislocation and the side surface.

In the next section, we present a tight-binding model on the cubic lattice
for a Weyl semimetal with a pair of Weyl nodes
at $\mib{k}_{\pm} = (0,0,\pm k_{0})$.
It is pointed out that no spontaneous charge current appears in the system
if the Fermi energy is located at the band center (i.e., $E_{F} = 0$).
In Sect.~3, we analyze the low-energy electron states in a cylindrical
Weyl semimetal by using a continuum approximation.
We clarify the behaviors of the 1D chiral states along a screw dislocation
and the 2D chiral states near a side surface.
We approximately determine the charge current
due to these chiral states at zero temperature.
In Sect.~4, we numerically obtain the spatial distribution of
the charge current at zero temperature
by using the model given in Sect.~2.
We treat two particular structures:
the system of a rectangular prism with two antiparallel screw dislocations
under the periodic boundary condition in the three directions
and that of a regular prism with a screw dislocation
under the open boundary condition in the three directions.
The last section is devoted to a summary.
We set $\hbar = 1$ throughout this paper.

\section{Model}

We introduce a tight-binding model for Weyl semimetals on a cubic lattice
with the lattice constant $a$, where lattice sites are specified by indices
$l$, $m$, and $n$, respectively, in the $x$-, $y$-, and $z$-directions.
The two-component state vector for the $(l,m,n)$th site is expressed as
\begin{align}
  |l,m,n \rangle
  =  \left[ |l,m,n \rangle_{\uparrow}, |l,m,n \rangle_{\downarrow} \right] ,
\end{align}
where $\uparrow, \downarrow$ represents the spin degree of freedom.
The tight-binding Hamiltonian is given by
$H = H_{0}+H_{x}+H_{y}+H_{z}$ with~\cite{yang,burkov1}
\begin{align}
   H_{0}
 & = \sum_{l,m,n} |l,m,n \rangle h_{0} \langle l,m,n| ,
         \\
   H_{x}
 & = \sum_{l,m,n}
     \left\{ |l+1,m,n \rangle h_{x} \langle l,m,n|
             + {\rm h.c.} \right\} ,
         \\
   H_{y}
 & = \sum_{l,m,n}
     \left\{ |l,m+1,n \rangle h_{y} \langle l,m,n|
             + {\rm h.c.} \right\} ,
         \\
   H_{z}
 & = \sum_{l,m,n}
     \left\{ |l,m,n+1 \rangle h_{z} \langle l,m,n|
             + {\rm h.c.} \right\} ,
\end{align}
where
\begin{align}
   h_{0}
 & = \left[ 
       \begin{array}{cc}
         2t\cos(k_{0}a) + 4B & 0 \\
         0 & -2t\cos(k_{0}a) - 4B
       \end{array}
     \right] ,
               \\
   h_{x}
 & = \left[ 
       \begin{array}{cc}
         -B & \frac{i}{2}A \\
         \frac{i}{2}A & B
       \end{array}
     \right] ,
               \\
   h_{y}
 & = \left[ 
       \begin{array}{cc}
         -B & \frac{1}{2}A \\
         -\frac{1}{2}A & B
       \end{array}
     \right] ,
               \\
   h_{z}
 & = \left[ 
       \begin{array}{cc}
         -t & 0 \\
         0 & t
       \end{array}
     \right] .
\end{align}
From the Fourier transform of $H$,
we find that the energy dispersion of this model is
\begin{align}
           \label{eq:exp-E}
   E =
 & \pm \Big\{\left[\Delta(k_{z})
                   +2B\bigl(2-\cos(k_{x}a)-\cos(k_{y}a)\bigr)\right]^{2}
              \nonumber \\
 & \hspace{10mm}
           + A^{2}\bigl(\sin^{2}(k_{x}a)+\sin^{2}(k_{y}a)\bigr)
       \Big\}^{\frac{1}{2}} ,
\end{align}
where
\begin{align}
  \Delta(k_{z}) = - 2t\left[\cos(k_{z}a)-\cos(k_{0}a)\right] .
\end{align}
A pair of Weyl nodes appears at $\mib{k}_{\pm} = (0,0,\pm k_{0})$ with $E = 0$
for moderate values of the parameters.
If the system is in the shape of a prism
with its top and bottom surfaces parallel to the $xy$-plane,
Fermi arc states appear only on the side surfaces.

Now, we introduce a screw dislocation parallel to the $z$-axis
in our tight-binding model without deforming the lattice structure itself.
We assume that the screw dislocation is centered at $(x_{\rm d},y_{\rm d})$
with $x_{\rm d}=(l_{\rm d}+\frac{1}{2})a$
and $y_{\rm d}=(m_{\rm d}+\frac{1}{2})a$,
and that it has a displacement of $N$ unit atomic layers
characterized by the Burgers vector $\mib{b} = a(0,0,N)$.
To take this into account, we consider a half plane (i.e., slip plane)
with its edge being identical to the dislocation line [see Fig.~2(a)]
and modify the hopping terms in $H$ across it so that each term connects
two different layers in the $z$-direction [see Fig.~2(c)].
As an example, let us consider the half plane parallel to
the $xz$-plane [i.e., $y = (m_{\rm d}+\frac{1}{2})a$].
In this case, we reconnect the hopping terms in $H_{y}$
for any $n$ in the region of $l_{\rm d}+1 \le l$
by performing the following replacement:
\begin{align}
 & |l,m_{\rm d}+1,n \rangle h_{y} \langle l,m_{\rm d},n|
             + {\rm h.c.}
       \nonumber \\
 & \to   
   |l,m_{\rm d}+1,n+N \rangle h_{y} \langle l,m_{\rm d},n|
             + {\rm h.c.} ,
\end{align}
which indicates that the site with $m = m_{\rm d}$ on the $n$th layer
is connected to the site with $m = m_{\rm d}+1$ on the $n+N$th layer
across the half plane.
Consequently, $H_{y}$ is modified as
\begin{align}
      \label{eq:mod-H_y}
   H_{y}
 & = \sum_{l_{\rm d}+1 \le l,m=m_{\rm d},n}
     \left\{ |l,m+1,n+N \rangle h_{y} \langle l,m,n|
             + {\rm h.c.} \right\} 
          \nonumber \\
 & + \sum_{\rm otherwise}
     \left\{ |l,m+1,n \rangle h_{y} \langle l,m,n|
             + {\rm h.c.} \right\}  .
\end{align}
If the system size $N_{z}$ in the $z$-direction is finite
(i.e., $1 \le n \le N_{z}$), as in realistic situations,
the summation over $n$ should be restricted within $1 \le n \le N_{z}-N$
in the first term in Eq.~(\ref{eq:mod-H_y}).
This reflects the fact that a straight step edge with height $N$ appears
on the top and bottom surfaces in the region of $l_{\rm d}+1 \le l$
[see Fig.~2(c)] owing to the presence of the screw dislocation.
\begin{figure}[btp]
\begin{center}
\includegraphics[height=7.0cm]{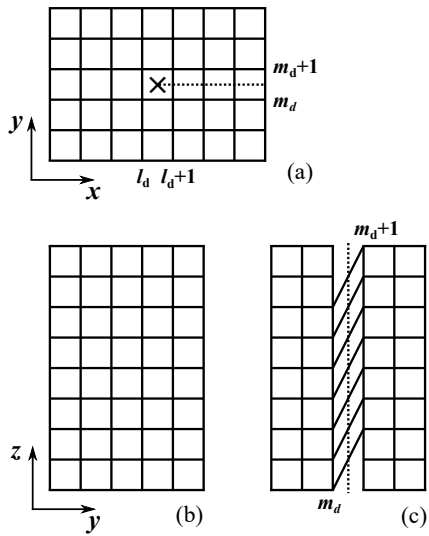}
\end{center}
\caption{
Lattice system in the presence of the screw dislocation with
a displacement of two unit atomic layers,
where each solid line between two neighboring sites represents
a hopping term in $H_{x}$, $H_{y}$, or $H_{z}$
that directly connects the two sites.
Dotted lines represent
the slip plane across which the hopping terms in $H$ are modified.
(a) Top view with the cross representing the dislocation center. 
(b) Side view in the unmodified region of $l \le l_{\rm d}$.
(c) Side view in the modified region of $l_{\rm d}+1 \le l$.
}
\end{figure}

We present the charge current operator $\mathcal{J}_{z}$ in the $z$-direction
on the link connecting the $(l,m,n)$th and $(l,m,n+1)$th sites.
It is given by
\begin{align}
  \mathcal{J}_{z}
   = iea\bigl[ |l,m,n+1\rangle h_{z} \langle l,m,n| - {\rm h.c.}
            \bigr] ,
\end{align}
which fully describes the charge current in the $z$-direction
in the absence of a screw dislocation parallel to the $z$-axis.
In the presence of such a screw dislocation, the tilted hopping term
[i.e., the first term of Eq.~(\ref{eq:mod-H_y})] also gives rise to
the charge current in the $z$-direction on the link connecting
the $(l,m_{\rm d},n)$th and $(l,m_{\rm d}+1,n+N)$th sites
only in the region of $l_{\rm d}+1 \le l$.
This contribution $\mathcal{J}_{z}^{\rm d}$ is expressed as
\begin{align}
  \mathcal{J}_{z}^{\rm d}
   = iea \bigl[ |l,m_{\rm d}+1,n+N\rangle h_{y} \langle l,m_{\rm d},n|
                  - {\rm h.c.} \bigr] .
\end{align}

Assuming the periodic boundary condition in the $z$-direction, we give
the energy band structures of the system in the two cases
considered in Sect.~4: the case with two antiparallel screw dislocations
piercing the rectangular cross section [Fig.~3(a)] and that with
one screw dislocation piercing the square cross section [Fig.~3(b)].
In the former case with two dislocations,
we impose the periodic boundary condition in the $x$- and $y$-directions
to eliminate the side surface, resulting in the absence of 2D chiral states.
In the latter case with a single dislocation,
the open boundary condition is imposed in the $x$- and $y$-directions;
thus, the 2D chiral states appear near the side surface.
The band structures in the two cases are shown in Figs.~4(a) and 4(b)
as a function of $k_{z}$, where the area of the cross section
$N_{x} \times N_{y}$ is $24 \times 12$ in the two-dislocation case
and $16 \times 16$ in the one-dislocation case.
The following parameters are used:
$B/A = t/A = 0.5$, $k_{0}a = 3\pi/4$, and $N = 2$.
Note that, as we show in the next section, the 1D chiral states appear
in the subgap region of $-k_{0} < k_{z} < k_{0}$
such that their branches cross the line of $E = 0$ at
\begin{align}
  q_{l} = \frac{2\pi}{Na}\left(-l-\frac{1}{2}\right)
\end{align}
satisfying
\begin{align}
      \label{eq:condition-1D}
     -k_{0} < q_{l} < k_{0} ,
\end{align}
where $l$ is an integer.
In the case of $N = 2$ with $k_{0}a = \frac{3\pi}{4}$,
a branch of the 1D chiral states crosses
the line of $E = 0$ at $k_{z}a = \frac{\pi}{2}$ or $-\frac{\pi}{2}$.
This is consistent with the band structures shown in Figs.~4(a) and 4(b).
As shown in Fig.~4(b), in the presence of the side surface, the subgap region
is filled with many branches with a nearly flat dispersion.
These branches represent the 2D chiral surface states.
Such branches disappear in the absence of the side surface
as clearly seen in Fig.~4(a).
\begin{figure}[btp]
\begin{center}
\includegraphics[height=2.0cm]{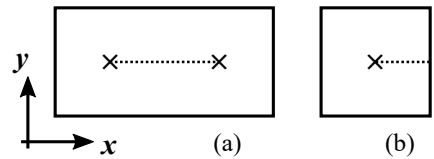}
\end{center}
\caption{
Cross sections of (a) the system with two antiparallel screw dislocations
and (b) that with one screw dislocation,
where dotted lines represent slip planes.
}
\end{figure}
\begin{figure}[btp]
\begin{center}
\includegraphics[height=4.0cm]{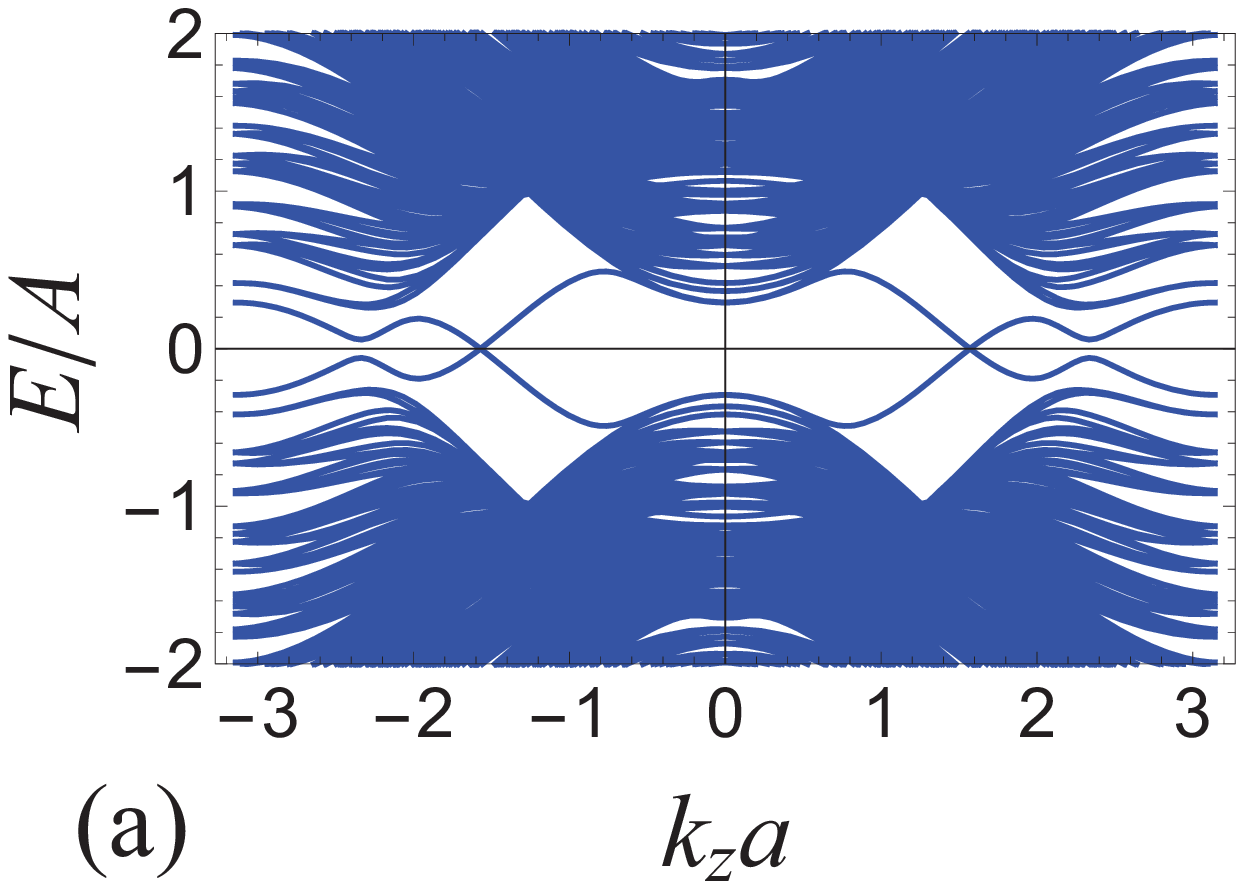}
\end{center}
\vspace{2mm}
\begin{center}
\includegraphics[height=4.0cm]{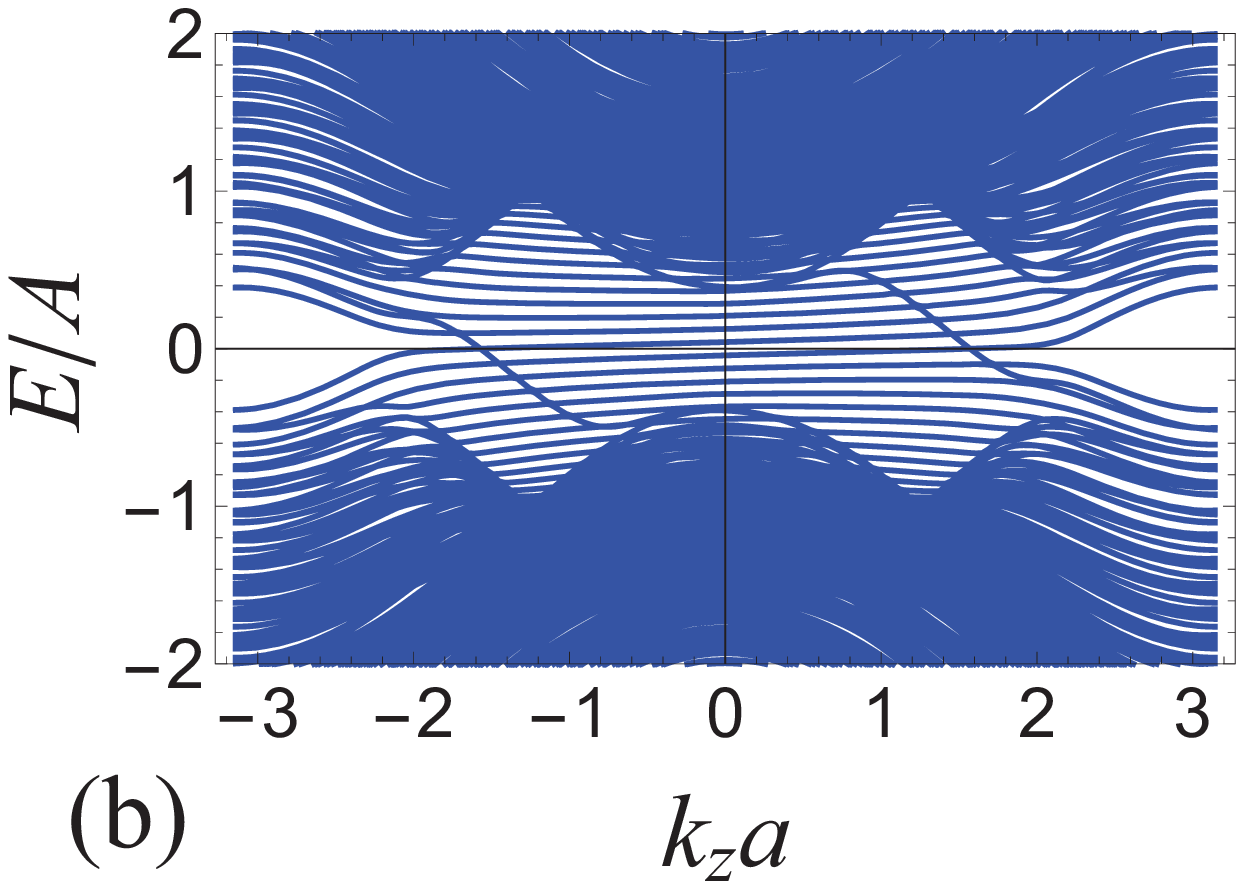}
\end{center}
\caption{
(Color online) Energy dispersions in the (a) two-dislocation case
with $N_{x} \times N_{y} = 24 \times 12$
and (b) one-dislocation case with $N_{x} \times N_{y} = 16 \times 16$.
}
\end{figure}

Here, it is important to point out that
if $E_{F}$ is located at the band center (i.e., $E_{F} = 0$),
no spontaneous charge current appears
in the system described by the model introduced above.~\cite{takane5}
This is directly related to the particle-hole symmetry characterized
by the operator $\Gamma_{\rm ph}$ defined by
\begin{align}
  \Gamma_{\rm ph} = \sigma_{x}K ,
\end{align}
where $\sigma_{x}$ and $K$ are respectively the $x$ component of
the Pauli matrices and the complex conjugate operator.
The tight-binding Hamiltonian $H$ satisfies
\begin{align}
    \label{eq:ph-symmet}
  \Gamma_{\rm ph}^{-1}H\Gamma_{\rm ph} = - H 
\end{align}
even in the presence of a screw dislocation.
By using this symmetry, we can verify that the local charge current completely
vanishes everywhere in the system at $E_{F} = 0$.~\cite{takane5}
This statement relies only on the particle-hole symmetry represented by
Eq.~(\ref{eq:ph-symmet}); thus, it is not restricted to the two-orbital model
used in this study and is also applicable to the four-orbital model
introduced in Ref.~\citen{vazifeh} and used in Ref.~\citen{sumiyoshi}.

\section{Analysis under a Continuum Approximation}

To estimate the magnitude of a charge current, we analyze the 1D chiral states
along a screw dislocation and the 2D chiral surface states
in a cylindrical Weyl semimetal of radius $R$ and length $L_{z}$.
In this analysis, a continuum approximation~\cite{takane1} is used
under the periodic boundary condition in the $z$-direction.
The Hamiltonian $H$ is rewritten in a continuum approximation as
\begin{align}
   H = \left[ 
         \begin{array}{cc}
           \Delta(k_{z})+B(\hat{k}_{x}^{2}+\hat{k}_{y}^{2})
              & A(\hat{k}_{x}-i\hat{k}_{y}) \\
           A(\hat{k}_{x}+i\hat{k}_{y})
              & -\Delta(k_{z})-B(\hat{k}_{x}^{2}+\hat{k}_{y}^{2})
         \end{array}
       \right] ,
\end{align}
where $\hat{k}_{x}=-i\partial_{x}$, $\hat{k}_{y}=-i\partial_{y}$.
By solving the eigenvalue equation
of $H\Psi = E\Psi$ with $\Psi(x,y) = {}^t\!(F,G)$,
we determine the energy dispersion relation of the 1D and 2D chiral states.

It is convenient to use the cylindrical coordinates $(r,\phi,z)$
with $r = \sqrt{x^{2}+y^{2}}$ and $\phi = \arctan(y/x)$.
We assume that a screw dislocation parallel to the $z$ axis
is inserted at $r = 0$ with a displacement of $N$ unit atomic layers.
We rewrite $F$ and $G$ as $F = e^{il\phi} f(r)$ and $G = e^{i(l+1)\phi} g(r)$,
where $l$ is the azimuthal quantum number.
The screw dislocation is described by
the effective vector potential:~\cite{kawamura}
$e(A_{r},A_{\phi},A_{z}) = (0,\zeta(k_{z})/r,0)$
with $\zeta(k_{z}) = \frac{Na}{2\pi}k_{z}$.
As demonstrated in Ref.~\citen{takane1}, the eigenvalue equation is
decomposed into the two sets of equations:
\begin{align}
        \label{eq:f-first}
   \left(\mathcal{D}_{l}-\Lambda_{-}\right)f = 0 ,
        \hspace{4mm}
   \left(\mathcal{D}_{l+1}-\Lambda_{-}\right)g = 0 
\end{align}
and
\begin{align}
        \label{eq:f-second}
   \left(\mathcal{D}_{l}-\Lambda_{+}\right)f = 0 ,
        \hspace{4mm}
   \left(\mathcal{D}_{l+1}-\Lambda_{+}\right)g = 0 ,
\end{align}
where 
\begin{align}
      \label{eq:def-D}
 \mathcal{D}_{l} = \partial_{r}^{2}+\frac{1}{r}\partial_{r}
                     - \frac{(l+\zeta)^{2}}{r^{2}} ,
\end{align}
and $\Lambda_{\pm}$ is given by
\begin{align}
  \Lambda_{\pm} = \frac{A^{2}+2B\Delta\pm\sqrt{(A^{2}+2B\Delta)^{2}
                        +4B^{2}(E^{2}-\Delta^{2})}}{2B^{2}} .
\end{align}
Assuming that $B$ is finite but very small, we can replace
$\Lambda_{-}$ and $\Lambda_{+}$ with
\begin{align}
          \label{eq:def-Lambda}
  \Lambda_{-} = -\frac{E^{2}-\Delta^{2}}{A^{2}} ,
         \hspace{6mm}
  \Lambda_{+} = \frac{A^{2}}{B^{2}} .
\end{align}
The two functions $f$ and $g$ are related by
the original eigenvalue equation with a very small $B$.
Hereafter, Eq.~(\ref{eq:f-first}) with $\Lambda_{-}$
and Eq.~(\ref{eq:f-second}) with $\Lambda_{+}$ are
respectively referred to as the Weyl equation and supplementary equation.
The set of two equations is much more tractable than
the original eigenvalue equation.

Let us express an eigenfunction for a given $l$ as $\Psi(r) = {}^t\!(f, g)$.
We require $\Psi(R) = {}^t\!(0,0)$ as a natural boundary condition.
In addition, $\Psi(0) = {}^t\!(0,0)$ is also required
in the presence of the screw dislocation at $r = 0$.
We focus on the case of $|\Delta|>|E|$, in which
the 1D and 2D chiral states appear.
In this case, the general solution $\Psi$ is written as
\begin{align}
  \Psi = c_{1} \psi_{l+\zeta}^{\eta} + d_{1} \psi_{l+\zeta}^{\kappa}
         + c_{2} \psi_{-l-\zeta}^{\eta}
         + d_{2} \psi_{-l-\zeta}^{\kappa} ,
\end{align}
where the functions with $\eta \equiv \sqrt{\Delta^{2}-E^{2}}/A$
are the solutions of the Weyl equation,
\begin{align}
    \psi_{l+\zeta}^{\eta}(r) & =
    {}^t\!\bigl[I_{l+\zeta}(\eta r), iR'(E)I_{l+1+\zeta}(\eta r)\bigr] ,
       \\
    \psi_{-l-\zeta}^{\eta}(r) & =
    {}^t\!\bigl[I_{-l-\zeta}(\eta r), iR'(E)I_{-l-1-\zeta}(\eta r)\bigr]
\end{align}
with $R'(E)=(-\Delta+E)/\sqrt{\Delta^{2}-E^{2}}$, and the functions
with $\kappa \equiv A/B$ are the solutions of the supplementary equation,
\begin{align}
   \psi_{l+\zeta}^{\kappa}(r) & =
   {}^t\!\bigl[I_{l+\zeta}(\kappa r), iI_{l+1+\zeta}(\kappa r)\bigr] ,
        \\
   \psi_{-l-\zeta}^{\kappa}(r) & =
   {}^t\!\bigl[I_{-l-\zeta}(\kappa r), iI_{-l-1-\zeta}(\kappa r)\bigr] .
\end{align}
Here, $I_{\nu}(x)$ is the $\nu$th-order modified Bessel function
of the first kind.
As modified Bessel functions and their linear combinations asymptotically
increase or decrease in an exponential manner,
$\Psi$ should describe chiral states spatially localized near $r = 0$ or $R$.
The energy dispersion relations of the chiral states are determined by
imposing the boundary conditions on $\Psi$.
See Ref.~\citen{takane1} for details of the derivation.

Now, we present the final result of energy dispersion relations
focusing on the low-energy regime of $E \approx 0$.
Both the 1D and 2D chiral states appear in the limited region of
$-k_{0} < k_{z} < k_{0}$, in which $\Delta(k_{z}) < 0$.
In this range of $k_{z}$, the low-energy 1D chiral states appear
near $k_{z} = q_{l}$ with
\begin{align}
      \label{eq;def-tilde_k_l}
  q_{l} = \frac{2\pi}{Na}\left(-l -\frac{1}{2} \right) .
\end{align}
The energy dispersion near $q_{l}$ is very steep and is given by
\begin{align}
       \label{eq:1D-chiral-dis}
  E_{\rm 1D}(k_{z},l)
     = -\frac{|\Delta(q_{l})|Na}{\pi}
       \ln\left(\frac{A^{2}}{B|\Delta(q_{l})|}\right)
       (k_{z}-q_{l}) .
\end{align}
This indicates that a branch of the 1D chiral states
appears for each $l$ satisfying Eq.~(\ref{eq:condition-1D}).
If $N = 2$ with $k_{0}a = \frac{3\pi}{4}$,
Eq.~(\ref{eq:condition-1D}) is satisfied
for $q_{0}a = -\frac{\pi}{2}$ and $q_{-1}a = \frac{\pi}{2}$,
indicating that the corresponding energy dispersions
respectively cross $E = 0$ at $k_{z} = q_{0}$ and $q_{-1}$.
If $N = 1$, no 1D chiral states appear irrespective of $k_{0}$.
In contrast to the 1D chiral states, the 2D chiral states have
a nearly flat dispersion as a function of $k_{z}$.
Near $E = 0$, their energy dispersion is given by
\begin{align}
       \label{eq:disp-CS_state}
  E_{\rm 2D}(k_{z},l) = \frac{ANa}{2\pi R}\left(k_{z}-q_{l}\right) ,
\end{align}
where the range of $k_{z}$ is limited by $-k_{0} < k_{z} < k_{0}$.
This clearly indicates that the energy dispersion becomes completely flat
in the absence of a screw dislocation (i.e., $N = 0$).

Now, we approximately determine the charge current $I_{z}$
in the $z$-direction at zero temperature,
assuming that $E_{F} > 0$ for simplicity.
As $I_{z}$ vanishes at $E_{F} = 0$,~\cite{comment} we are allowed to pick up
only the contribution from the energy range of $0 < E < E_{F}$.
Let us determine $I_{z}^{\rm 1D}$ due to the 1D chiral states.
The group velocity is
\begin{align}
  v_{\rm 1D} = -\frac{|\Delta(q_{l})|Na}{\pi}
            \ln\left(\frac{A^{2}}{B|\Delta(q_{l})|}\right) ,
\end{align}
and the number of $k_{z}$ satisfying $0 < E_{\rm 1D}(k_{z},l) < E_{F}$
is given by
\begin{align}
   \text{Number of $k_{z}$}
   = \frac{E_{F}L_{z}}
          {2Na|\Delta(q_{l})|
            \ln\left(\frac{A^{2}}{B|\Delta(q_{l})|}\right)}
\end{align}
for each $l$ satisfying $-k_{0} < q_{l} < k_{0}$.
Thus, the charge current due to the 1D chiral states is given by
\begin{align}
      \label{eq:1D-result}
    \frac{I_{z}^{\rm 1D}}{L_{z}} =  e \frac{n_{\rm 1D}E_{F}}{2\pi} ,
\end{align}
where $n_{\rm 1D}$ represents the number of allowed $l$
(i.e., the number of branches corresponding to the 1D chiral states).
For example, $n_{\rm 1D} = 2$
in the case of $N = 2$ and $k_{0}a = \frac{3\pi}{4}$.
We turn to the contribution from the 2D chiral states.
The group velocity in the $z$-direction is
\begin{align}
  v_{\rm 2D} = \frac{ANa}{2\pi R} ,
\end{align}
and the numbers of $k_{z}$ and $l$ satisfying
$0 < E_{\rm 2D}(k_{z},l) < E_{F}$ are approximately given by
\begin{align}
 &  \text{Number of $k_{z}$}
   = \frac{k_{0}L_{z}}{\pi}
     \\
 &   \text{Number of $l$} = \frac{RE_{F}}{A} .
\end{align}
Thus, the charge current due to the 2D chiral states is obtained as
\begin{align}
      \label{eq:2D-result}
    \frac{I_{z}^{\rm 2D}}{L_{z}} =  -e \frac{N(k_{0}a)E_{F}}{2\pi^{2}} .
\end{align}
Equations~(\ref{eq:1D-result}) and (\ref{eq:2D-result}) hold
even in the case of $E_{\rm F} < 0$.
Although $I_{z}^{\rm 1D}$ and $I_{z}^{\rm 2D}$ are determined under
the assumption of $B$ being very small,
$B$ disappears in their final expressions.
We thus expect that the resulting $I_{z}^{\rm 1D}$ and $I_{z}^{\rm 2D}$ are
reliable irrespective of the value of $B$.
In the case of $N = 2$ and $k_{0}a = \frac{3\pi}{4}$, we find that
\begin{align}
       \label{eq:Iz-N_2}
 &  \frac{I_{z}^{\rm 1D}}{L_{z}}
    =  e \frac{E_{F}}{\pi} ,
        \\
 &  \frac{I_{z}^{\rm 2D}}{L_{z}}
    =  -e \left(\frac{3}{4}\right)\frac{E_{F}}{\pi} .
\end{align}
Generally, $I_{z}^{\rm 1D}$ and $I_{z}^{\rm 2D}$ flow in the opposite direction
from each other and their magnitude is not identical, leading to
the conclusion that the addition of $I_{z}^{\rm 1D}$ and $I_{z}^{\rm 2D}$
does not necessarily vanish even after being integrated over a cross section.
This means that the bulk states compensate the difference between them.

\section{Numerical Results}

\begin{figure}[btp]
\begin{center}
\includegraphics[height=3.7cm]{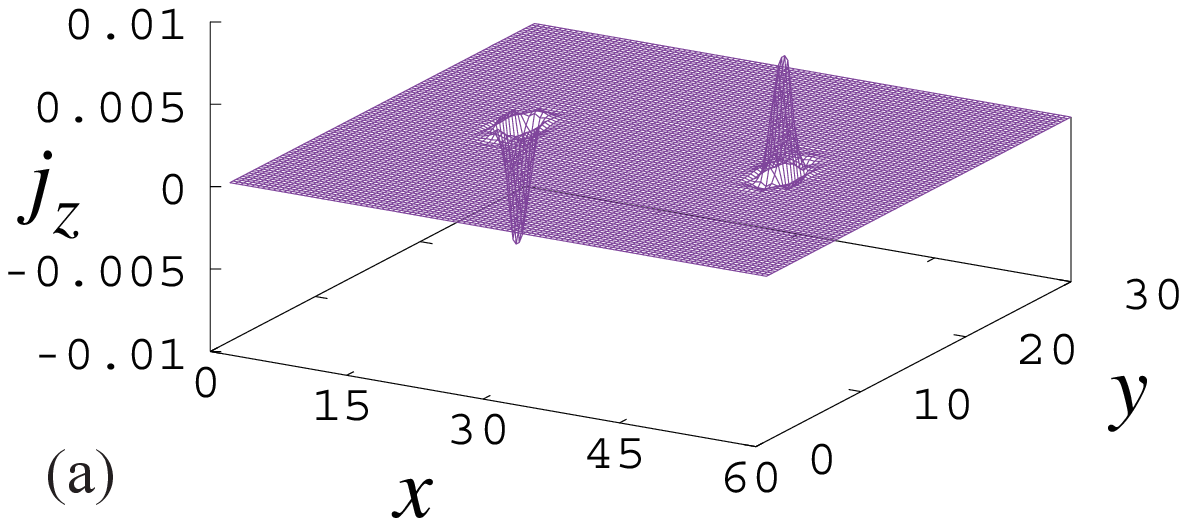}
\end{center}
\vspace{4mm}
\begin{center}
\includegraphics[height=3.7cm]{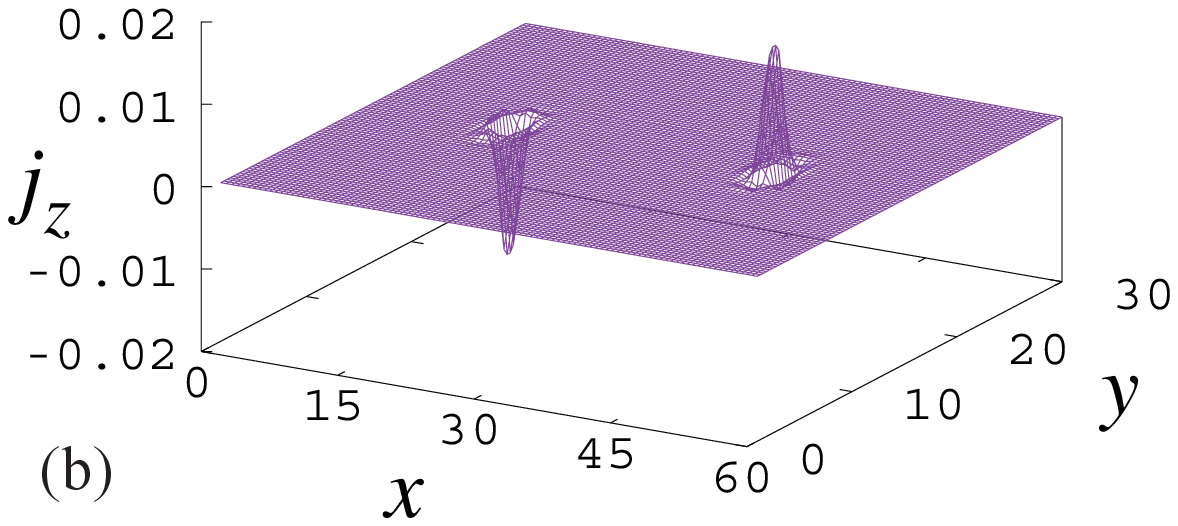}
\end{center}
\vspace{4mm}
\begin{center}
\includegraphics[height=3.7cm]{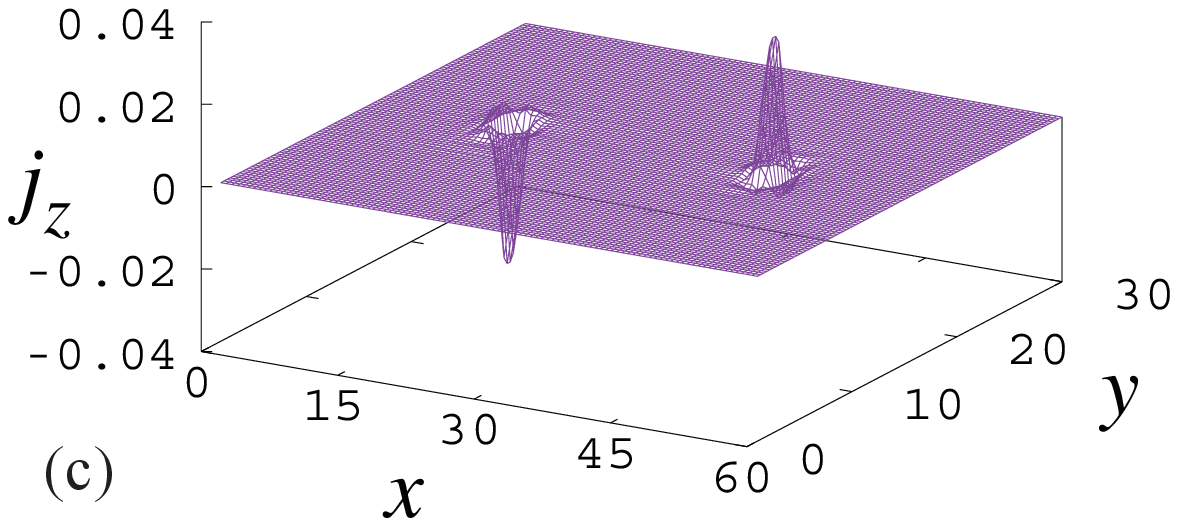}
\end{center}
\vspace{4mm}
\begin{center}
\includegraphics[height=3.7cm]{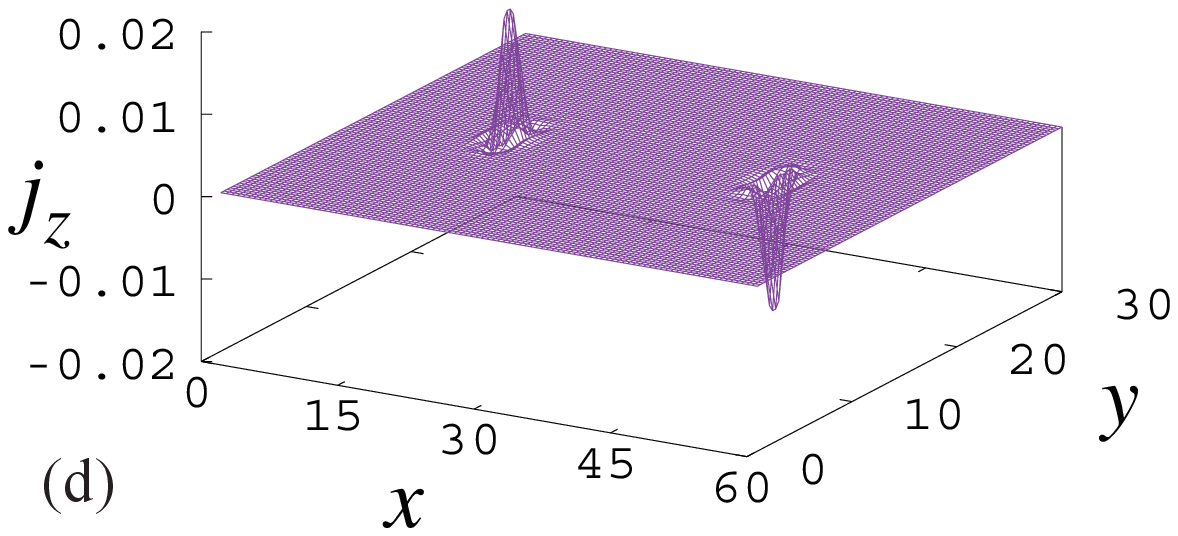}
\end{center}
\caption{
(Color online) $j_{z}$ normalized by $eA N_{z}a$ in the two-dislocation case
with $N = 2$ on the cross section parallel to the $xy$-plane.
(a) $E_{\rm F}/A = 0.05$, (b) $0.1$, (c) $0.2$, and (d) $-0.1$.
Note that the vertical scale is different from figure to figure.
}
\end{figure}
In this section, we give the numerical results of the local charge current
$j_{z}$ in the $z$-direction and $j_{x}$ in the $x$-direction
at zero temperature.
They are obtained by calculating the average of the corresponding operators
over the eigenstates with an energy smaller than $E_{\rm F}$.
In the two-dislocation case [see Fig.~3(a)], we give $j_{z}$
on the cross section parallel to the $xy$-plane to clarify
the effect of the 1D chiral states on the persistent current.
In the one-dislocation case [see Fig.~3(b)], we give $j_{z}$ and $j_{x}$
on a few cross sections to observe the spatial distribution of
the persistent current in a closed system.

Figure~5 shows $j_{z}$ on the cross section parallel to the $xy$-plane
in the two-dislocation case with $N = 2$, where
the system of $N_{x} \times N_{y} \times N_{z} = 60 \times 30 \times 240$
is used in calculating $j_{z}$
under the periodic boundary condition in the three directions.
In this figure, $j_{z}$ is normalized by $eA N_{z}a$.
The Fermi energy is located at $E_{\rm F}/A = 0.05$, $0.1$, $0.2$, and $-0.1$,
where the result at $E_{\rm F} = 0$ is not shown
since the local charge current vanishes everywhere in the system.
This is consistent with the argument given in Sect.~2.
From Figs.~5(a)--5(c), we clearly observe that the magnitude of $j_{z}$
is nearly proportional to $E_{\rm F}$, which is consistent with
Eq.~(\ref{eq:1D-result}) given in Sect.~3.
We also observe that $j_{z}$ changes its sign between
the result at $E_{\rm F}/A = 0.1$ [Fig.~5(b)]
and that at $E_{\rm F}/A = -0.1$ [Fig.~5(d)]
without varying its magnitude.
This is also consistent with Eq.~(\ref{eq:1D-result}).
As $j_{z}$ is positive in the region of $31 \le x \le 60$,
the charge current $I_{z}$ in the $z$-direction is obtained by integrating
$j_{z}$ in this region.
At $E_{\rm F}/A = 0.1$, the result is
\begin{align}
  \frac{I_{z}}{L_{z}} \approx 0.019 \times eA
\end{align}
with $L_{z} = N_{z}a$.
This is roughly consistent with Eq.~(\ref{eq:Iz-N_2}).
Figure~6 shows $j_{z}$ in the two-dislocation case with $N = 1$
at $E_{\rm F}/A = 0.1$.
The 1D chiral states are absent in this case.
As clearly seen from Fig.~(6), the magnitude of $j_{z}$ is much smaller
than that in the case with $N = 2$ at $E_{\rm F}/A = 0.1$ [Fig.~5(b)].
This clearly indicates that the 1D chiral states crucially contribute
to the persistent current.
\begin{figure}[btp]
\begin{center}
\includegraphics[height=3.7cm]{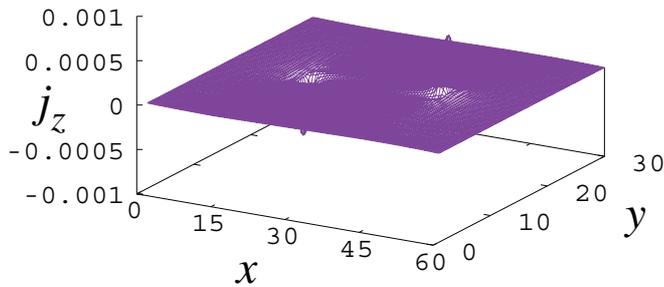}
\end{center}
\caption{
(Color online) $j_{z}$ normalized by $eA N_{z}a$ in the two-dislocation case
with $N = 1$ on the cross section parallel to the $xy$-plane
at $E_{\rm F}/A = 0.1$.
}
\end{figure}

We now give the numerical results of $j_{z}$ and $j_{x}$
in the one-dislocation case with $N = 2$
on the three cross sections indicated in Fig.~7.
In the figures given below, $j_{z}$ and $j_{x}$ are normalized by $eAa$.
The open boundary condition is imposed on the system of
$N_{x}\times N_{y}\times N_{z} = 40 \times 40 \times 60$
in the three directions
with the Fermi energy fixed at $E_{\rm F}/A = 0.1$.
Figure~8(a) shows $j_{z}$ on the cross section parallel to the $xy$-plane
denoted by the dashed (red) line in Fig.~7.
The peak structure at the center of the cross section represents
the charge current in the $z$-direction due to the screw dislocation.
In addition, the charge current in the opposite direction appears
near the crossing point between the slip plane and the side surface
at $x = 40$ and $y = 20$.
This contribution arises from the 2D chiral states
circulating around the side surface in a spiral manner.
It appears only near the crossing point simply because
the screw dislocation is included in our model by modifying
only the hopping terms across the slip plane [see Fig.~2(c)].
Figures~8(b) and 8(c) show $j_{x}$ on the cross sections parallel
to the $yz$ plane denoted by the short-dash-dotted (blue) and
long-dash-dotted (green) lines in Fig.~7.
In both figures, $j_{x}$ flows in the $-x$-direction ($x$-direction)
near the line edge located at $y = 1$ ($y = 40$).
This represents the charge current due to the 2D chiral states,
which circulates around the side surface.
In Fig.~8(b), a pair of peak structures is observed near the points
at $y = 20$ on the bottom surface (i.e., $z = 1$)
and on the top surface (i.e., $z = 60$).
As a step edge is located at these points,
the two peaks should be identified as the charge current
along the edge dislocations on the top and bottom surfaces.
Such peak structures do not appear in Fig.~8(b), reflecting the fact that
no edge dislocation is present on the corresponding cross section
denoted by the long-dash-dotted (green) line in Fig.~7.
Figures~8(a)--8(c) indicate that the charge current
due to the 1D chiral states is converted to the spiral surface current
due to the 2D chiral states
mainly through the edge dislocations located at the top and bottom surfaces.
\begin{figure}[btp]
\begin{center}
\includegraphics[height=4.0cm]{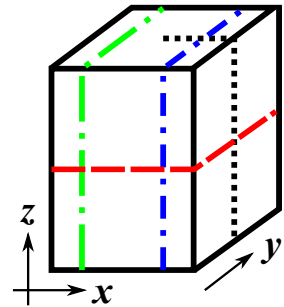}
\end{center}
\caption{
(Color online) System in a rectangular parallelepiped shape
with a screw dislocation in the $z$-direction,
where the dotted line represents slip plane.
In the text, the charge current is considered
on the three cross sections denoted by
dashed (red), short-dash-dotted (blue), and long-dash-dotted (green) lines.
}
\end{figure}
\begin{figure}[btp]
\begin{center}
\includegraphics[height=3.7cm]{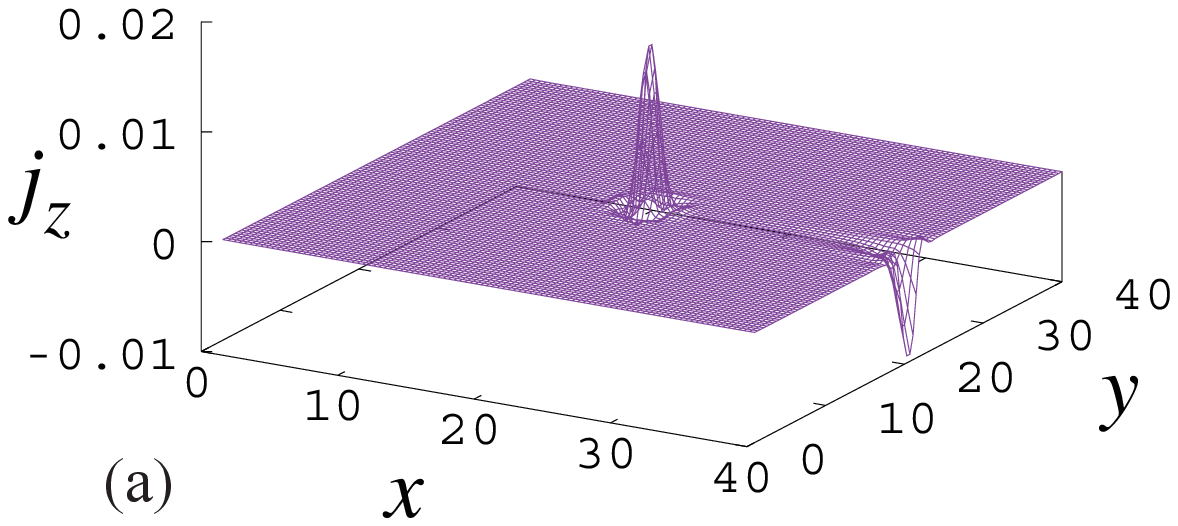}
\end{center}
\vspace{4mm}
\begin{center}
\includegraphics[height=3.7cm]{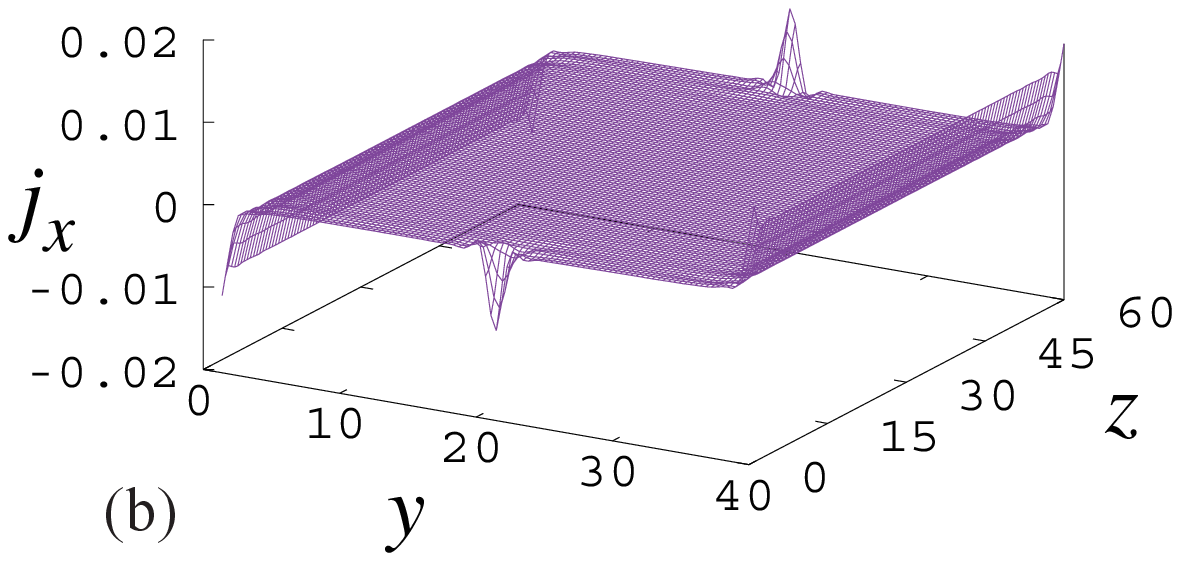}
\end{center}
\vspace{4mm}
\begin{center}
\includegraphics[height=3.7cm]{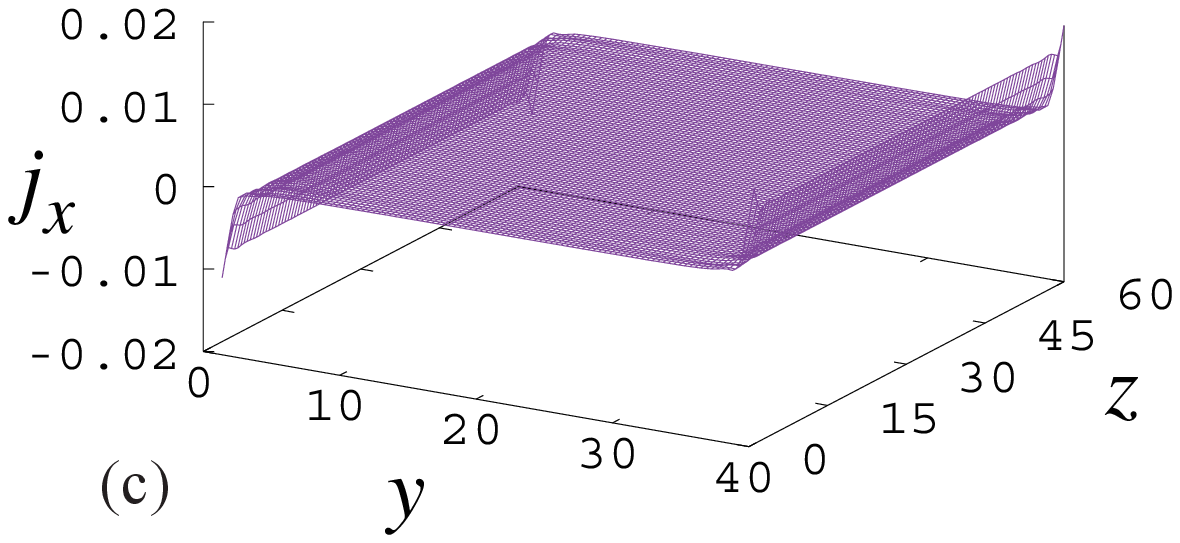}
\end{center}
\caption{
(Color online) $j_{z}$ and $j_{x}$ normalized by $eAa$
in the one-dislocation case with $N = 2$ at $E_{\rm F}/A = 0.1$
on the three cross sections denoted in Fig.~7:
(a) $j_{z}$ on the cross section denoted by the dashed (red) line,
(b) $j_{x}$ on the cross section denoted by the short-dash-dotted (blue) line,
and (c) $j_{x}$ on the cross section denoted by the long-dash-dotted (green)
line.
}
\end{figure}

\section{Summary}

We analyzed the local charge current in a Weyl semimetal in the presence of
a screw dislocation focusing on the role of 1D chiral states
that appear along the corresponding dislocation line.
We found several interesting features
that were not observed in the previous study.~\cite{sumiyoshi}
The most important feature is that the contribution to the charge current
from the 1D chiral states significantly dominates that from the bulk states.
Another feature is that the local charge current significantly depends on
the location of the Fermi energy $E_{F}$.
If $E_{F}$ is at the band center,
the local charge current vanishes everywhere in the system.
A finite charge current appears near the dislocation
when $E_{F}$ is displaced from the band center.
The magnitude of the local charge current is nearly proportional to $E_{F}$
as long as $E_{F}$ is located near the band center.
We also considered how the charge current due to the 1D chiral states is
converted to the surface charge current, as well as the bulk charge current,
at the top and bottom surfaces of the system.
It is shown that the conversion is mainly mediated through
the edge dislocation,
which connects the screw dislocation and the side surface.

\section*{Acknowledgment}

This work was supported by JSPS KAKENHI Grant Number JP18K03460.


\begin{thebibliography}{99}
\bibitem{comment} This fact is not taken into consideration
in Refs.~\citen{takane1} and \citen{sumiyoshi}.
\end{thebibliography}
\end{document}